\documentclass[twocolumn, 11pt, a4paper]{article}
\usepackage[absolute]{textpos}
\usepackage[top=3cm,bottom=3cm,left=2cm,right=2cm]{geometry}
\usepackage{amsmath,amssymb,mathtools,dsfont,emptypage,graphicx,dcolumn,bm,booktabs,colortbl,collcell,enumerate,float,verbatim,multirow,xparse,slashed,mathrsfs,color}
\usepackage{cite}
\usepackage{tabularx}
\usepackage{multirow}
\usepackage{graphicx}
\usepackage{subcaption}
\usepackage{appendix}
\usepackage{booktabs} 
\usepackage{colortbl}
\usepackage{collcell}
\usepackage{xparse}
\usepackage{relsize}
\usepackage{multicol}
\usepackage{titlesec}
\usepackage{float}
\usepackage{ragged2e}
\usepackage[labelfont=bf, labelsep=period]{caption}
\usepackage[compat=1.1.0]{tikz-feynman}
\usepackage{contour}
\usepackage{footmisc}
\usepackage{scrextend}
\deffootnote[1em]{1.5em}{1em}{\textsuperscript{\thefootnotemark}\,}
\usepackage{etoolbox} 
\titleformat{\section}
  {\normalfont\large\bfseries} 
  {\thesection. }{.1em}{}         

\titleformat{\subsection}
  {\normalfont\normalsize\bfseries}
  {\thesubsection. }{.1em}{}

\titleformat{\subsubsection}
  {\normalfont\normalsize\bfseries}
  {\thesubsubsection. }{.1em}{}

\BeforeBeginEnvironment{pmatrix}{
  \setlength{\arraycolsep}{1.5pt} 
}
\AfterEndEnvironment{pmatrix}{
  \setlength{\arraycolsep}{5pt} 
}

\usepackage[compat=1.1.0]{tikz-feynman}
\usepackage{contour}

\newcommand{\virg}[1]{``#1''}

\def\Javg{\mathcal{J}_{\rm avg}}
\def\DM{\mathsmaller{\rm DM}}
\def\FO{\mathsmaller{\rm FO}}

\definecolor{myorange}{rgb}{0.8, 0.3, 0.2}
\definecolor{myblue}{rgb}{0.27, 0.5, 0.9}
\definecolor{mygray}{gray}{.95}
\usepackage[colorlinks=true,urlcolor=myorange,linkcolor=myorange,citecolor=myorange]{hyperref}

\begin{document}

\twocolumn[{%
\begin{@twocolumnfalse}
\begin{center}
{\bf \large Dark Matter Interpretation of the Super-Kamiokande  Antineutrino Excess\\ and Predictions for JUNO} \\
[5mm]
\renewcommand*{\thefootnote}{\fnsymbol{footnote}}
Alessandro Granelli$^{\,a,}$\hyperlink{email1}{${}^\dagger$},
Silvia Pascoli$^{\,b,c,}$\hyperlink{email2}{${}^\ddagger$} and
Salvador Rosauro-Alcaraz$^{\,d,}$\hyperlink{email3}{${}^\S$}
\\
\vspace{2mm}
$^{a}${\it Instituto de Física Corpuscular (IFIC), CSIC-Universitat de València, Parc Científic UV,\\ c/ Catedrático José Beltrán 2, E-46980 Paterna, Spain}\\
$^{b}${\it Dipartimento di Fisica e Astronomia, Università di Bologna, via Irnerio 46, 40126, Italy} \\
$^{c}${\it INFN, Sezione di Bologna, viale Berti Pichat 6/2, 40127, Bologna, Italy}\\
$^{d}${\it Université Clermont-Auvergne, CNRS, LPCA, 63000 Clermont-Ferrand, France,\\
*LPCA pour Laboratoire de Physique de Clermont Auvergne}
\end{center}
\begin{center}
    {\bf Abstract}
\end{center}
Super-Kamiokande has reported a small excess of electron antineutrino events in the 20 MeV energy range, in the search for the diffuse supernova neutrino background. We interpret this signal as a possible indication of dark matter that annihilates dominantly into neutrinos, pointing to a thermal dark matter candidate with $s$-wave annihilation and with mass in the tens of MeV range. This mass scale naturally fits into rich dark sector extensions of the Standard Model. Neutrino experiments, including JUNO, will be able to test this hypothesis in the coming years. \\
\hrule
\vspace{1em}
  \end{@twocolumnfalse}
}]

\begingroup
\renewcommand{\thefootnote}{}
\footnotetext{%
\hspace{.45em}\hypertarget{email1}{${}^\dagger$} \href{mailto:alessandro.granelli@ific.uv.es}{alessandro.granelli@ific.uv.es}\\
\hypertarget{email2}{${}^\ddagger$} \href{mailto:silvia.pascoli@unibo.it}{silvia.pascoli@unibo.it}\\
\hypertarget{email3}{${}^\S$} \href{mailto:salvador.rosauro@clermont.in2p3.fr}{salvador.rosauro@clermont.in2p3.fr}
}
\endgroup

\renewcommand*{\thefootnote}{\arabic{footnote}}
\setcounter{footnote}{0}

\textbf{\textit{Introduction ---}}~After operating for over two decades with high-purity water, the Super-Kamiokande (SK) detector \cite{Suzuki:2019jby} has recently been loaded with gadolinium(III) sulfate octohydrate, $\text{Gd}_2(\text{SO}_4)_3\cdot\,8\text{H}_2\text{O}$ 
\cite{Super-Kamiokande:2021the, Super-Kamiokande:2024kcb}. As first suggested in \cite{Beacom:2003nk}, owing to Gd's exceptionally high neutron-capture cross-section and the $\sim 8\,\text{MeV}$ $\gamma$-rays following its de-excitation, Gd-loaded water-Čerenkov detectors have an enhanced capability to tag interactions producing final-state neutrons by identifying the neutron-capture $\gamma$-rays in (delayed) coincidence with the prompt event signal. For this reason, Gd-loaded SK (SK-Gd) has significantly increased the sensitivity to, e.g., 
cosmic-ray spallations, atmospheric and reactor neutrinos, and the Diffuse Supernovae Neutrino Background (DSNB).

The latter is the integrated neutrino flux originated from all the past distant supernovae \cite{Seidov1982:sva, Krauss:1983zn, Dar:1985mm}. Recent predictions for this flux are in the range $ \sim 1$--$6  
\ \mathrm{cm^2 \ s^{-1}}$ corresponding to up to few events per year in SK, see e.g.~\cite{Ekanger:2023qzw, Ando:2023fcc,Volpe:2023met,Santos:2025inv}. Large uncertainties still affect the predictions, the dominant ones being due to the fraction of failed supernovae forming black holes, the neutrino emission spectra at late times and the cosmic star formation rate.

The DSNB has not been detected yet, but it has been a target of intensive searches for over a decade: first via positron tagging \cite{Super-Kamiokande:2002hei, Super-Kamiokande:2011lwo},
then neutron-capture on protons \cite{Super-Kamiokande:2013ufi, Super-Kamiokande:2021jaq},
and later neutron-capture on Gd \cite{Super-Kamiokande:2023xup, Super-Kamiokande:2025sxh}.
Intriguingly, an excess of electron antineutrino  ($\bar \nu_e$) events in the energy range 
relevant to DSNB searches, specifically around an antineutrino energy $E_{\bar \nu} \approx 22\,\text{MeV}$, has been reported at SK, already at $\sim 1.5\sigma$ when operating with pure water \cite{Super-Kamiokande:2021jaq} and later after the second Gd-loading \cite{Super-Kamiokande:2025sxh} (see also \cite{Harada:2024DSNB}). When interpreted as a DSNB signal and considering only post-Gd-loadings data \cite{Super-Kamiokande:2025sxh}, the background-only hypothesis is disfavoured at $0.5\sigma - 1.7\sigma$, depending on the adopted DSNB model \cite{Totani:1995rg, Hartmann:1997qe, Malaney:1996ar, Kaplinghat:1999xi, Fukugita:2002qw, Horiuchi:2008jz, Lunardini:2009ya, Galais:2009wi, Nakazato:2015rya, Priya:2017bmm, Barranco:2017lug, Horiuchi:2017qja, DeGouvea:2020ang, Horiuchi:2020jnc, Kresse:2020nto, Tabrizi:2020vmo, Ashida:2023heb, Ivanez-Ballesteros:2022szu, Martinez-Mirave:2024zck, Nakazato:2024gem} and neutron tagging approach, and up to $\sim 2.3\sigma$ when all SK runs are combined \cite{Harada:2024DSNB} (see also \cite{Beauchene:2024DSNB,  Rogly:2024DSNB, Santos:2024SKGd}). The sizeable theoretical uncertainty in the prediction of DSNB flux spans over more than one order of magnitude, with certain models predicting a $\bar{\nu}_e$ flux at $\sim 22\,\text{MeV}$ ranging from about $10\%$ to $50\%$ of the flux required to account for the excess \cite{Hartmann:1997qe, Malaney:1996ar, Nakazato:2015rya, Barranco:2017lug, Ivanez-Ballesteros:2022szu} (see also Figs.~1 and 12 of Ref.~\cite{Super-Kamiokande:2025sxh}). This renders a DSNB interpretation not yet unambiguous and leaves room for alternative explanations.

In this letter, we explore the intriguing possibility that the SK $\bar \nu_e$ excess is an indication of sub-GeV dark matter (DM). The existence of DM in the Universe is well established observationally (see e.g.~\cite{Cirelli:2024ssz} for a comprehensive review). 
In \cite{Palomares-Ruiz:2007trf} it was pointed out that self-annihilations of local MeV-scale DM particles into neutrinos 
would be a smoking-gun signature at neutrino detectors in the same energy ballpark of DSNB searches. The first bounds on the self-annihilation cross-section for DM mass $m_{\DM}$ in the range $15 \ \mathrm{MeV} \lesssim m_{\DM} \lesssim 130 \ \mathrm{MeV} $ were obtained \cite{Palomares-Ruiz:2007trf} using the SK positron events \cite{Super-Kamiokande:2002hei}. It was found that, depending on the local DM profile assumed, cross-sections as small as the one required to produce the DM relic abundance via thermal freeze-out
could be tested. Subsequent analyses have refined this bound using data from the first pure-water SK runs \cite{Primulando:2017kxf, Olivares-DelCampo:2017feq, Arguelles:2019ouk}.

Here, we analyse the SK data from all six runs, including those after the Gd-loadings, and evaluate the statistical significance of the excess when interpreted as a DM signal. We consider both the case in which DM annihilates directly into neutrinos, leading to a very peaked signature~\cite{Palomares-Ruiz:2007trf}, or into dark particles that subsequently decay into neutrinos with a typically broader spectrum. We take into account
the improved knowledge of the astrophysical parameters describing the local DM density that makes the interpretation of the reported $\bar \nu_e$ excess easier to accommodate as thermal sub-GeV DM annihilations, compared to previous analysis.

Importantly, new experiments will be able to test our hypothesis. The JUNO experiment~\cite{JUNO:2015zny, JUNO:2022lpc}, currently taking data, as well as Hyper-Kamiokande (HK)~\cite{Hyper-Kamiokande:2018ofw}, will have sensitivity in the mass range of interest. We present the predictions in JUNO: thanks to its excellent energy resolution, it will be able to distinguish between a peaked signal, as predicted in our case, and the broad one from the DSNB.\\

\noindent \textbf{\textit{Dark matter annihilations into neutrinos ---}} 
In the standard scenario of thermal freeze-out, 
the DM relic abundance is controlled by the total annihilation cross-section $\sigma_{\rm tot}$, in particular by $\langle \sigma_{\rm tot} v \rangle_{\FO}$ with $v$ the DM relative velocity and $\langle \cdot \rangle_{\FO}$ the thermal average computed at freeze-out. Explaining the value of $\Omega_{\DM}h^2 = 0.12$ as measured by PLANCK \cite{Planck:2018vyg} with $s$-wave annihilations
 requires $\langle\sigma_{\rm tot} v \rangle_{\FO}\simeq 3.8\,(4.3) \times 10^{-26} \text{cm}^3 \,\text{s}^{-1}$ for $m_{\DM} = 10\,(100)\,\text{MeV}$ for a real scalar/Majorana DM -- 
somewhat larger than the canonical $3 \times 10^{-26} \mathrm{cm}^3 \mathrm{s}^{-1}$, see e.g.~\cite{Steigman:2012nb}. These values would be double in the case of complex scalar/Dirac DM. 

In the DM mass range of interest $m_{\DM} \sim (10- 100)\,\text{MeV}$, annihilations into charged particles or photons cannot saturate the value required \cite{Essig:2013goa,Slatyer:2015jla,Boudaud:2016mos,Ng:2019gch,Laha:2020ivk,Cirelli:2023tnx,Siegert:2024hmr,Wang:2025tdx} and annihilations uniquely into light stable or long-lived dark sector particles are also disfavoured since the latter would contribute to the effective number of relativistic degrees of freedom affecting Big Bang Nucleosynthesis (BBN), see e.g. \cite{Ho:2012ug,Nollett:2014lwa,Sabti:2019mhn,Sabti:2021reh, Yeh:2026pil}. An alternative is to have DM annihilations into dark sector particles that promptly decay dominantly into neutrinos, directly or via decay chains, as typical in rich dark sectors~\cite{Ballett:2019pyw, Abdullahi:2025fiy}. As per the case of direct annihilations above, decays into charged particles, photons or stable/longlived light dark sector particles are constrained to be subdominant~\cite{Elor:2015bho}. Therefore, annihilation into neutrinos is 
the Standard Model (SM) channel of choice, the one that can (directly or indirectly) dominate DM annihilations and lead the generation of the present DM relic abundance. 

DM-neutrino interactions have been studied both from a theoretical point of view \cite{Batell:2017cmf,Blennow:2019fhy}, often linking it to neutrino mass generation at the sub-GeV scale~\cite{Boehm:2006mi,Arhrib:2015dez,Li:2022bpp, Xu:2023xva}, and for their cosmological consequences~\cite{Boehm:2014vja,Olivares-DelCampo:2017feq,Akita:2023yga,Brax:2023tvn,Brinckmann:2022ajr,Heston:2024ljf,Zu:2025lrk, Kanemura:2025byi, Dev:2025tdv}. Annihilations $\text{DM} \ \text{DM} \rightarrow \bar{\nu} \nu$ can proceed via the exchange of dark sector mediators. Typically, SM gauge invariance can be guaranteed by invoking e.g.~mixing between neutrinos and heavy neutral leptons (HNLs). This also implies that annihilations into charged leptons are subdominant but thermal annihilation cross-sections generally
require large mixing angles \cite{Blennow:2019fhy}. In the case of annihilation into dark sector particles, all couplings involved refer to the dark sector and can be sizable so that sufficiently large cross-sections can be easily achieved. Their decays, e.g.~dark scalar or HNL decays into neutrinos, can be sufficiently fast even in the presence of small mixing, intriguingly compatible with those required for neutrino masses~\cite{RDSDM} and with BBN constraints.

For concreteness, we consider two options to explain the SK $\bar{\nu}_e$ excess: DM annihilating directly into a neutrino-antineutrino pair, or DM annihilating into a pair of dark particles, say two real scalars $\phi$, each promptly decaying into a neutrino-antineutrino pair. We note that other possibilities are viable, for instance DM annihilating into one neutrino and one HNL or into two HNLs. The latter subsequently decay, e.g., into neutrinos and/or other dark sector particles, with possibly complex decay chains and with a rather broad neutrino spectrum. We leave the exploration of these further possibilities for a future related work.\\ 

\noindent \textbf{\textit{The antineutrino flux ---}}  The differential diffuse 
antineutrino flux (per-flavour) from non-relativistic DM annihilations is given by \cite{Palomares-Ruiz:2007trf, Olivares-DelCampo:2017feq} (see also, e.g., \cite{Arguelles:2019ouk}):
\begin{equation}
\frac{d\Phi_{\bar \nu}}{dE_{\bar \nu}} = \frac{\left\langle\sigma v \right\rangle_0}{4\pi \kappa m_{\DM}^2} \frac{1}{3}\frac{dN_{\bar \nu}}{dE_\nu} J,
\label{eq:general_nu_flux}
\end{equation}
where $\kappa=2\,(4)$ if the DM particle is a real (complex) scalar/Majorana (Dirac) fermion; $\langle\sigma v \rangle_0$ is the annihilation cross-section times relative velocity, thermally averaged over the present DM velocity distribution; the factor of $1/3$ accounts for flavour equilibration due to neutrino oscillations; $dN_{\bar \nu}/dE_{\bar \nu}$ is the number of antineutrinos emitted per unit of energy per annihilation, whose expression depends on the antineutrino production chain; the $J$-factor is the line-of-sight (LOS) integral of the square of the local DM density $\rho_{\DM}$. 

In the case of DM directly annihilating into a neutrino-antineutrino pair, the antineutrino spectrum is monochromatic, reading 
$dN_{\bar\nu}/dE_{\bar\nu}=\delta(E_{\bar\nu}-m_{\DM})$.  
In the case of DM annihilating into two scalars first, the formal expression for the flux remains unchanged, provided that $\langle \sigma v \rangle_0$ is understood as the annihilation cross-section into scalars, and $dN_{\bar\nu}/dE_{\bar\nu}$ is doubled and uniform, with an average energy of $m_{\DM}/2$ and a width of $\delta E_{\bar \nu}=m_{\DM}
\sqrt{2\Delta} \sqrt{1-\Delta/2}$, where $\Delta \equiv (m_{\DM}-m_\phi)/m_{\DM}$ with $0<\Delta\leq 1$, and $m_\phi$ is the mass of the scalar. 

The $J$-factor in Eq.~\eqref{eq:general_nu_flux} is defined as
\begin{equation}
J \equiv \int_{\Delta \Omega} d\Omega\int_0^{\ell_{\rm max}} d\ell\, \rho_{\DM}^2(r(\ell,\mu)),
\end{equation}
where $r(\ell, \mu) = \sqrt{R_0^2 + \ell^2 - 2 \ell R_0 \mu}$; $R_0$ 
is the galactocentric distance; $\mu \equiv \cos \alpha$, $\alpha \in [0, \pi]$ being the angle between the LOS $\ell$ and the Sun-Galactic Center (GC) direction; $d\Omega = d\varphi\,d\mu$, $\varphi \in [0, 2\pi]$ being the azimuthal angle and $\Delta \Omega$ the total solid angle of view. For a spherical DM halo of radius $R_{\rm halo}$ 
centred in the GC, the maximal LOS reads $\ell_{\rm max} = \sqrt{R_{\rm halo}^2 -R_0^2 (1-\mu^2) } + R_0 \mu $.  
The galactocentric distance $R_0$ has been relatively well determined by the GRAVITY collaboration, finding $R_0 \simeq 8.277\,\text{kpc}$ \cite{GRAVITY:2021xju} (see also \cite{Abuter:2021yys}). The integral does not depend much on $R_{\rm halo}$ as long as this is sufficiently large. For definiteness, we consider $R_{\rm halo} = 100\,\text{kpc}$ in our estimations. We are interested in full-sky signals, and thus set $\Delta \Omega = 4\pi$.
Furthermore, we find it practical to normalise the $J$-factor to $J_0 =  4\pi R_0\rho_0^2$, with the DM density at $R_0$, $\rho_0 \equiv \rho_{\DM}(R_0)$, fixed at the reference value $0.4\, \text{GeV}\,\text{cm}^{-3}$, and express our results in terms of the dimensionless \virg{averaged} quantity $\Javg\equiv J/J_0$. 

The limited knowledge on the exact shape of the local DM profile implies a relatively large uncertainty in the determination of the $J$-factor.  A widely adopted parameterisation for the local DM distribution is the generalised version of the Navarro-Frenk-White (gNFW) profile \cite{Navarro:1995iw} 
\begin{equation}
    \rho_{\DM}(r) =  \rho_s \left(\frac{r}{r_s}\right)^{-\gamma} \left(1+\frac{r}{r_s}\right)^{\gamma-3},
\end{equation}
where $r_s$ is the scale radius and $\rho_s = 2^{3-\gamma} \rho_{\DM}(r_s)$ is the scale density. This form is motivated by N-body simulations, which predict a universal slope of $\gamma = 1$ for typical galaxies. Leaving $\gamma$ free as in the gNFW allows to account for baryonic effects and other uncertainties in the inner halo. The gNFW profile has been adopted, e.g., in \cite{Benito:2019ngh} (see also \cite{Benito:2020lgu}) to fit the Milky Way rotation curve data from stellar tracers in circular orbit up to $\sim 20\,\text{kpc}$. By using the likelihood provided in \cite{Benito:2019ngh} for $R_0 = 8.3\,\text{kpc}$ and profiling over $\gamma$, $\rho_s$ and $r_s$ in the ranges $1\leq \gamma\leq 1.5$, $0\leq \rho_s\leq 1.96\,\text{GeV}\,\text{cm}^{-3}$ and $5\leq r_s/\text{kpc}\leq 100$, we obtain a $\text{best-fit}\,\pm\,1\sigma$ value of $\mathcal{J}_{\rm avg} = 6^{+3}_{-1}$ and  $4\,(3) \lesssim \mathcal{J}_{\rm avg} \lesssim 17\,(30)$ at $2\sigma$ ($3\sigma$), reflecting the relatively large astrophysical uncertainties.\footnote{The same analysis yields the following $\text{best-fit}\,\pm\,3\sigma$ value for the local DM density at $R_0$: $\rho_0= 0.66^{+0.16}_{-0.20}\,\text{GeV}\,\text{cm}^{-3}$.} To compute the aforementioned ranges, we truncated the angular integration at $\alpha = 0.01^\circ$, thus effectively ignoring regions around the GC below $\sim 1\,\text{pc}$.\\

\noindent \textbf{\textit{Analysis of Super-Kamiokande data ---}}~The SK Collaboration has so far performed six analysis in search for the DSNB, accumulating more than 6000 days of data taking. They can be organised into the ``SK pure-water era'' (SK-I to SK-IV), the ``SK neutron-tagging era'' (SK-IV to SK-VII), and the ``SK-Gd era'' (SK-VI and SK-VII). Each one brought improvements on the analysis allowing to reduce backgrounds and thus have better sensitivity to the DSNB signal arising when an electron antineutrino interacts through inverse beta decay (IBD) in the SK detector. Such analyses can be used to search for DM annihilation signals in SK~\cite{Palomares-Ruiz:2007trf,Olivares-DelCampo:2017feq,Arguelles:2019ouk}. The expected signal in SK in the $i$-th energy bin is given by
\begin{equation}
	\begin{split}
	A_i=&N_{\mathrm{tar}} T\int_{E_{\mathrm{vis}}^i}^{E_{\mathrm{vis}}^{i+1}}dE_{\mathrm{vis}}\,\epsilon(E_{\mathrm{vis}}) \,\times \\
	\int& dE_e R(E_e,E_{\mathrm{vis}})\int_{E_{\bar{\nu}}^{\mathrm{min}}}dE_{\bar{\nu}}\frac{d\sigma^{\mathrm{IBD}}}{dE_e}\frac{d\Phi_{\bar{\nu}}}{dE_{\bar{\nu}}},
	\end{split}
	\label{eq:number_of_events}
\end{equation}
where $N_{\mathrm{tar}}$ is the number of protons in the $22.5$~kton fiducial mass of the SK detector; $T$ corresponds to the running time of the given SK analysis; $\epsilon$ stands for the efficiency, which is a function of the reconstructed positron energy $E_\mathrm{vis}$. We impose the same lower limit on the neutrino energy as done in \cite{Super-Kamiokande:2025sxh}, with $E_{\bar{\nu}}^{\mathrm{min}}=17.3$~MeV. Finally, $d\sigma^{\mathrm{IBD}}/dE_e$ is the differential IBD cross-section which we take from \cite{Strumia:2003zx}, and $R(E_e,E_\mathrm{vis})$ corresponds to the energy smearing for the positrons. In the following, we assume that the latter can be described as a Gaussian 
with width~\cite{Palomares-Ruiz:2007trf,Super-Kamiokande:2010tar,Super-Kamiokande:2023jbt} $\sigma_{\mathrm{res}}=0.40\sqrt{E_e/\textrm{MeV}}+0.03 E_e /\textrm{MeV}$ for every SK run, with the exception of SK-II. For the latter, given the reduction of the PMT coverage, 
we instead consider~\cite{Super-Kamiokande:2008ecj} $\sigma_{\mathrm{res}} = 0.0536 + 0.52 \sqrt{E_e / \textrm{MeV}} + 0.0458  E_e / \textrm{MeV}$.
We note that, even though the energy resolution has somewhat improved since SK-I, our choice for SK-III to SK-VII is conservative for the energy region we are interested in.

We perform a $\chi^2$-analysis of all SK data in order to assess the sensitivity to DM annihilations using the data from \cite{Super-Kamiokande:2002hei,Super-Kamiokande:2011lwo,Super-Kamiokande:2013ufi,Super-Kamiokande:2021jaq,Super-Kamiokande:2023xup,Super-Kamiokande:2025sxh,Beauchene:2024DSNB,Rogly:2024DSNB,Santos:2024SKGd,Harada:2024DSNB}. For each SK run, we define:
\begin{align}
\chi^2(\alpha,\vec{\theta})=-2\sum_{i=1}^{N_{\mathrm{bins}}}&\Big\lbrace O_i-\left[\alpha A_i+(1+\theta_i) B_i\right]\\
	+& O_i \log{\frac{\alpha A_i + (1+\theta_i) B_i}{O_i}}\Big\rbrace+\frac{\theta_i^2}{\sigma_i^2}\,,\nonumber
	\label{eq:chi^2}
\end{align}
where $O_i$ are the observed number of events, $B_i$ are the background events, $\theta_i$ are systematic uncertainties which we introduce as nuisance parameters with Gaussian priors of width $\sigma_i$, and $\alpha$ is the unknown normalization of the signal. We assume that the backgrounds are the same ones as in the DSNB, and marginalise over the nuisance parameters $\theta_i$. The value of $\sigma_i$ in each energy bin is fixed in order to reproduce previous SK results on the sensitivity to the DSNB~\cite{Super-Kamiokande:2025sxh,Beauchene:2024DSNB,Rogly:2024DSNB,Santos:2024SKGd,Harada:2024DSNB}, after which we only exchange the input DSNB signal for the DM annihilation one.

\begin{figure}[t!]
    \includegraphics[width=0.99\linewidth]{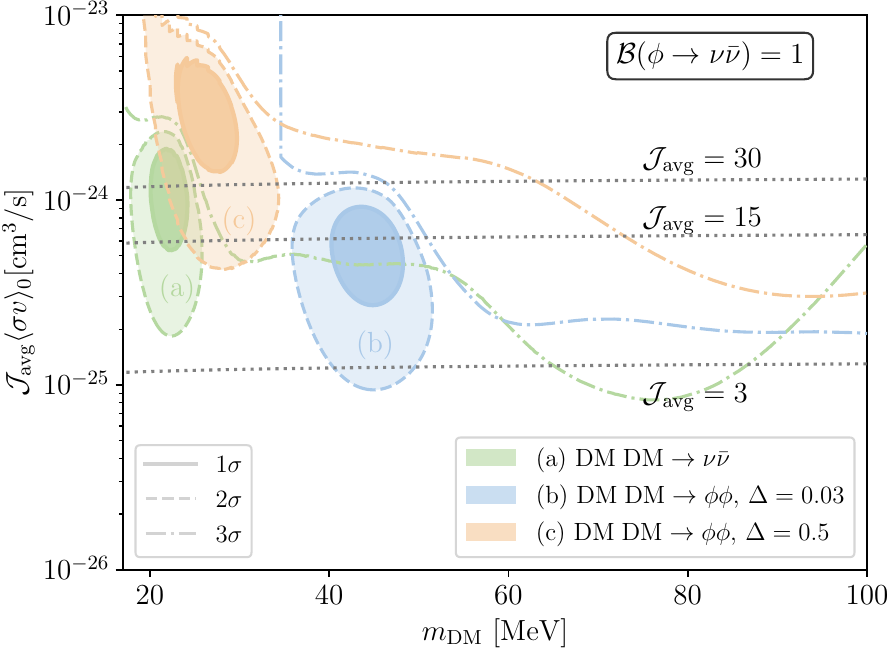}
    \caption{Preferred regions for $m_{\DM}$ and $\mathcal{J}_{\mathrm{avg}}\langle \sigma v\rangle_0$ for (a) $\text{DM DM} \to \nu {\bar \nu}$ (green), and $\text{DM DM} \to \phi \phi \to \nu \nu {\bar \nu} {\bar \nu} $ for (b) $\Delta = 0.03$ (blue) and (c) $\Delta = 0.5$ (orange), combining all SK runs. We assume that the scalar $\phi$ decays only into neutrinos. The solid, dashed and dot-dashed contours correspond to the $\Delta \chi^2$ levels associated to the $1\sigma$, $2\sigma$ and $3\sigma$ confidence regions, respectively. The horizontal dotted lines correspond to benchmark values for $\mathcal{J}_{\mathrm{avg}}$, for which the value of $\langle \sigma v\rangle_0$ equals the thermal relic target for s-wave annihilation $\langle \sigma v\rangle_{\mathrm{\FO}}$ \cite{Steigman:2012nb}. }
\label{fig:1}
\end{figure}

In Fig.~\ref{fig:1}, we present the results of the analysis. For direct DM annihilations (green), the data prefer a DM mass of $m_{\DM}^{\text{best-fit}} = 22.1\,\text{MeV}$ at the best fit point, and $\mathcal{J}_{\text{avg}} \langle \sigma v \rangle_0 \in [1.8 \times 10^{-25} $, $2.3 \times 10^{-24}] \ \text{cm}^3 \,\text{s}^{-1}$ at 2$\sigma$. Taking into account the allowed values of $\mathcal{J}_{\text{avg}}$, the annihilation cross section into neutrinos is compatible with that required for thermal dark matter freeze-out, $\langle\sigma_{\rm tot} v \rangle_{\FO}\simeq 4 \times 10^{-26} \text{cm}^3 \,\text{s}^{-1}$.  If DM annihilates into scalars that subsequently decay into neutrinos, the preferred mass depends on the width of the neutrino spectrum, which is controlled by the mass difference between DM and the scalar. We show the results for two benchmark choices for the scalar-DM mass splitting: $\Delta = 0.03$ (blue) and $\Delta = 0.5$ (orange). For sufficiently small mass splitting, the signal resembles that of the direct annihilation, albeit with a factor of two in the flux. For the benchmark $\Delta = 0.03$, we find $m_{\DM}^{\text{best-fit}} = 44.4\,\text{MeV}$ 
 and $\mathcal{J}_{\text{avg}} \langle \sigma v \rangle_0 \in [9.1 \times 10^{-26} $, $1.1 \times 10^{-24}] \ \text{cm}^3\, \text{s}^{-1}$ at 2$\sigma$, again compatible with the thermal freeze-out cross-section. For $\Delta = 0.5$, the spectrum is much broader providing a worse fit to the data and lowering the preferred masses to the $(20-30)\,\text{MeV}$ mass range. Higher
values of $\mathcal{J}_{\text{avg}} \langle \sigma v\rangle$ the cross-section are required in this case, still allowing consistency with the thermal freeze-out one but at a lower statistical significance.

The DM annihilation signal has a significance of $2.57\sigma$ for direct annihilation, $2.59\sigma$ for indirect annihilation with nearly degenerate DM and scalar ($\Delta = 0.03$), and $2.45\sigma$ for indirect annihilation with larger mass splitting ($\Delta = 0.5$), to be compared to the SK DSNB case. For the case of ${\rm DM}\, {\rm DM} \to \nu \bar{\nu}$ neutrinos (${\rm DM}\, {\rm DM} \to \phi\phi \to \nu\nu \bar{\nu}\bar{\nu}$), we find a $\chi^2_{\rm min} = 104.9\,(104.8)$ for a total of $179$
degrees of freedom (d.o.f). For the DSNB hypothesis we find instead $\chi^2_{\rm min} = 108.2$ with $180$ d.o.f.~The DM signal thus improves the fit by $\Delta \chi^2 \simeq 3.3$ relative to the DSNB hypothesis, with one additional
parameter.

\begin{figure}[t!]
\centering
    \includegraphics[width=0.95\linewidth]{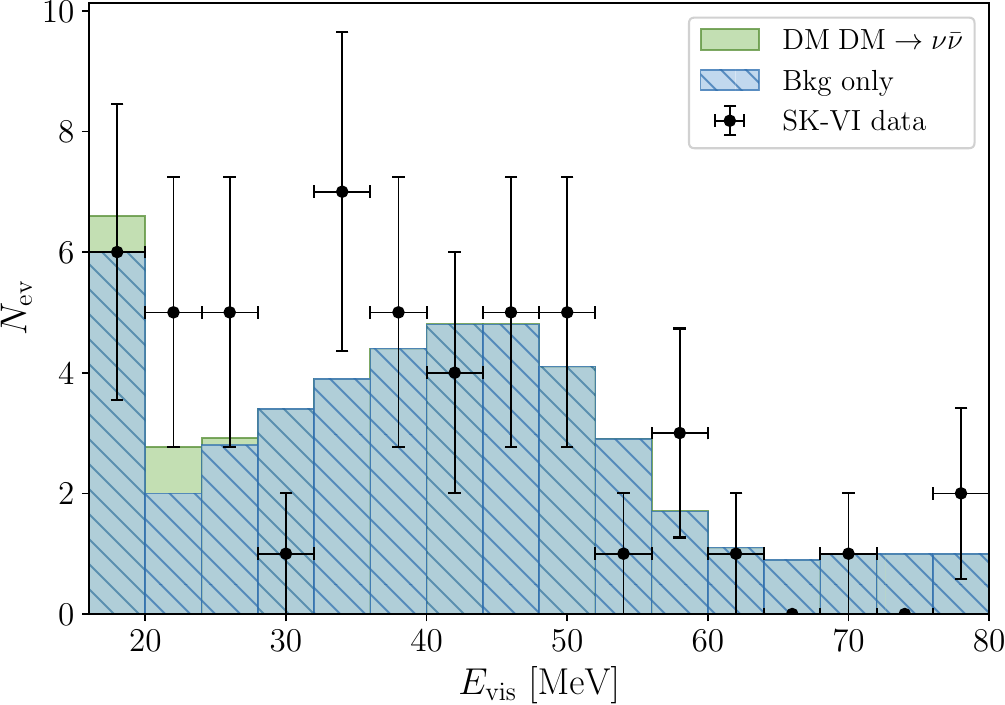}
    \includegraphics[width=0.99\linewidth]{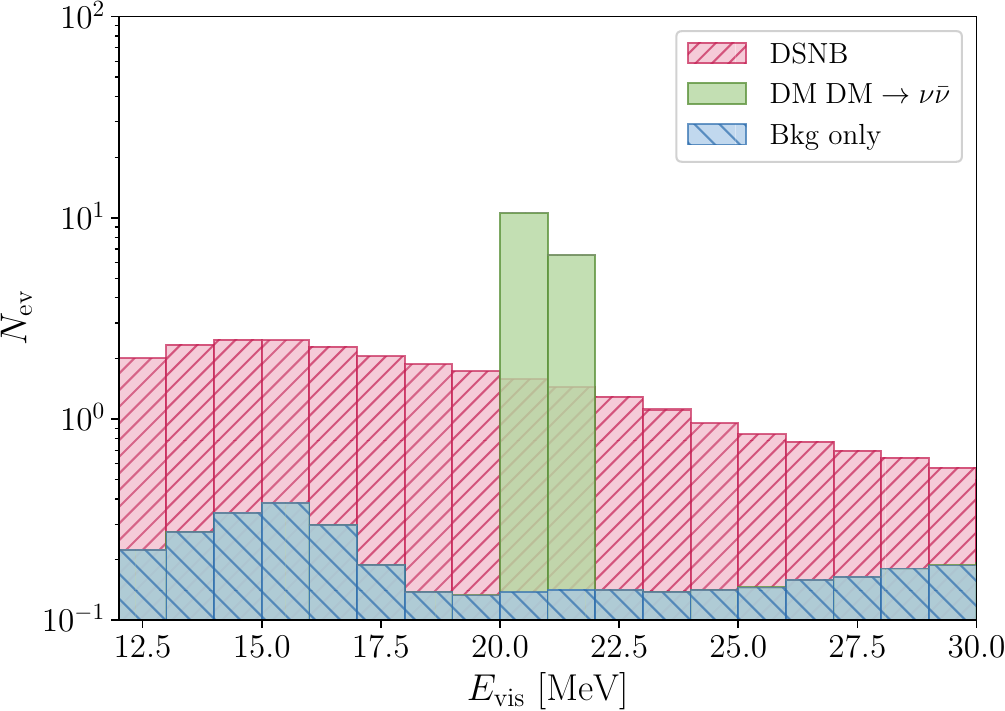}
    \caption{Predicted number of events in SK-VI (top) and JUNO (bottom) as a function of the reconstructed positron energy. For JUNO we considered a 147 $\mathrm{kton}\times \mathrm{yr}$ exposure. The black points are the data from SK-VI~\cite{Super-Kamiokande:2025sxh} with the respective statistical error bars. In blue we show the expected backgrounds (those for JUNO are taken from~\cite{JUNO:2022lpc}), while in green the expectation from direct DM annihilations into neutrinos that fit the best SK data. In the lower panel, we show in pink the event distribution for the DSNB that could also fit the SK data.}
\label{fig:2}
\end{figure}

We stress that these results are obtained under the assumption that the DSNB flux falls within the most conservative set of expected scenarios. Certain theoretical models predict a $\bar{\nu}_e$ flux at $22\,\text{MeV}$ corresponding to about $10\%\sim 50\%$ of that required to explain the excess \cite{Hartmann:1997qe, Malaney:1996ar, Nakazato:2015rya, Barranco:2017lug, Ivanez-Ballesteros:2022szu}, which we considered negligible for the purpose of our analysis. If instead the analysis is performed by including a DSNB flux as an additional background with a free normalisation, a similar best-fit point is obtained, but an open $2\sigma$ contour. In this case, the significance is reduced by almost 6 $\chi^2$-units.\footnote{Including the DSNB with its normalisation fixed at the SK best-fit value leads to
little room for any additional signal, and both the significance of the DM signal and its preferred normalisation are drastically reduced.}

Additional SK runs as well as current and upcoming neutrino experiments, such as HK~\cite{Hyper-Kamiokande:2018ofw} and JUNO\cite{JUNO:2015zny, JUNO:2022lpc}, will be able to test this hypothesis in the coming future. In particular, JUNO is a 20~kton LSc detector ideally suited to detect IBD, with an exceptional energy resolution of 3\%~\cite{JUNO:2022lpc}. The latter property is crucial for testing our hypothesis since the DM and the DSNB signals have different energy features. For illustration, we show in the top panel of Fig.~\ref{fig:2} the predicted number of events in SK-VI, and in the bottom panel the corresponding event rates in JUNO for $147\,\text{kton} \times \text{yr}$, using the best-fit results we obtained for the direct annihilation case (shown in green) and for the DSNB (shown in pink).
We find that the peaked DM signal clearly stands above both the background and the possible DSNB contribution, and would thus be readily distinguishable.\\

\noindent \textbf{\textit{Discussion and Conclusions ---}}
We have proposed an explanation for the SK electron-antineutrino excess around 20 MeV as a signal of DM (direct or indirect) annihilations into neutrinos. 
We find that the excess can be interpreted as an indication of a direct $\text{DM DM} \rightarrow \bar{\nu} \nu$ annihilation for a DM mass $m_{\DM}$ of around 22.1~MeV, with an annihilation cross-section compatible with thermal freeze-out for allowed local DM profiles. 
This is the simplest case, pointing to DM annihilating primarily into neutrinos via $s$-wave. Other possibilities could also be considered. Depending on the value of $\mathcal{J}_{\text{avg}}$, if the annihilation cross-section is higher than the freeze-out one or if it is velocity-dependent, a non-thermal production of DM at temperatures below its mass needs to be invoked. If the cross-section is instead smaller, other dominant annihilation channels need to be considered at freeze-out, for instance into dark sector particles that subsequently decay into lower-energy neutrinos.

For indirect annihilation into neutrinos via intermediate real scalars,
and a sufficiently small DM-scalar mass splitting, the signal has a very similar shape as for direct annihilation, but requires $m_{\DM}\sim 44.4\,\text{MeV}$ and $\mathcal{J}_\text{avg} \langle \sigma v \rangle$ lower approximately by a factor of 2. For larger mass splitting, e.g.~$0.5 \,m_{\DM}$, the spectrum is broader, pointing to the $(20-30)\,\text{MeV}$ mass range and higher values of the annihilation cross-section. For this case as well, the reconstructed annihilation cross-section would be compatible with the $s$-wave one responsible for DM freeze-out and similar considerations for larger and lower values could be made as for the direct annihilation case.

We point out a possible intriguing coincidence with the 511 keV line due to positron annihilations in the bulge. Despite being of long standing, this anomaly has not yet received a convincing astrophysical explanation and DM annihilations into electron-positron pairs have been advocated as a possible source~\cite{Boehm:2003bt,Ascasibar:2005rw,laTorreLuquePedro:2024est}. Recent analyses suggest that a spiky profile can induce the observed flux for DM masses up to 20 MeV without contradicting constraints from disk emission and in-flight positron annihilation with the interstellar medium, and even for higher injection positron energy~\cite{Das:2025tdh}. Notice that the branching ratio into electron-positrons is required to be much smaller than unity, compatible with our requirement of dominant annihilation into neutrinos. In the case of DM annihilations directly into neutrinos and into electron-positrons, the DM mass implied by the SK signal, 22.1~MeV, is in the range required for the 511 keV line~\cite{laTorreLuquePedro:2024est}. In the case of annihilation first into dark scalars, it should be taken into account that the energy injection of the electron/positrons from the scalar decays are much lower, making the higher DM mass again compatible with the 511~keV signal.

Moreover, we note that the energy scale invoked by the signal is typical of rich dark sectors~\cite{Abdullahi:2025fiy} and might provide additional connections with the generation of neutrino masses and, e.g., a supercooled first order phase transition implied by Pulsar-Timing-Array gravitational wave observatories~\cite{NANOGrav:2023hvm}.

This signal might possibly be the first glimpse of sub-GeV DM in particle physics experiments. Future evidence for an excess of events over an extended energy region in SK would disfavour the hypothesis of direct annihilation in favour of the continuous SM DSNB. JUNO and HK have the potential to further test our hypothesis in the forthcoming future. Thanks to their larger mass and better energy resolution, they may be able to distinguish a DM annihilation signal from that from DSNB and provide information on the DM properties. Time will tell.

\section*{Acknowledgements}

We thank Michele Lucente, Sergio Palomares-Ruiz and Filippo Sala for useful discussions. 
We also thank Elina Merkel for presenting a preliminary result from this work at the joint PLANCK2026 and 6th EuCAPT Symposium. 
We acknowledge the use of computational resources from the parallel computing cluster of the \href{https://site.unibo.it/openphysicshub/en}{Open Physics Hub} at the Department of Physics and Astronomy of the University of Bologna, as well as the Tier-3 HPC Cluster from the INFN Sezione di Bologna. The work of A.G.~is supported by the Spanish grant PID2023-148162NB-C21 (MCIN/AEI/10.13039/501100011033) co-funded by the European Union (FEDER), and in part by the  
European Union's Horizon Europe programme under the Marie Skłodowska-Curie Actions – Staff Exchanges (SE) grant agreement No.~101086085-ASYMMETRY. The work of S.P.~has been partly funded by the European Union under the Horizon Europe’s Project: 101201278 – DarkSHunt - ERC - 2024 ADG. Views and opinions expressed are however those of the author(s) only and do not necessarily reflect those of the European Union or the European Research Council Executive Agency. Neither the European Union nor the granting authority can be held responsible for them.


\providecommand{\href}[2]{#2}\begingroup\raggedright\endgroup

\end{document}